\documentclass[11pt]{article}

\usepackage{fancyhdr}
\usepackage{balance}
\usepackage{graphicx}
\usepackage{acronym}
\usepackage{amsmath} 
\usepackage{subfigure}
\newcommand{\comments}[1]{} 
\usepackage{float}
\usepackage{url}

\newcounter{MYtempeqncnt}

\newcommand{\ie}{\emph{i.e.}} 
\newcommand{\eg}{\emph{e.g.}}

\acrodef{OPEX}[OPEX]{Operating Expenses}
\acrodef{UE}[UE]{User Equipment}
\acrodef{BS}[BS]{Base Station}
\acrodef{DTX}[DTX]{Discontinuous Transmission}
\acrodef{PAPR}[PAPR]{Peak-to-Average Power Ratio }
\acrodef{SC-FDMA}[SC-FDMA]{Single-carrier FDMA}
\acrodef{FDMA}[FDMA]{Frequency Division Multiple Access}
\acrodef{TDMA}[TDMA]{Time Division Multiple Access}
\acrodef{CDMA}[CDMA]{Code Division Multiple Access}
\acrodef{OFDMA}[OFDMA]{Orthogonal Frequency Division Multiple Access}
\acrodef{ICT}[ICT]{Information and Communication Technologies}
\acrodef{QoS}[QoS]{Quality of Service}
\acrodef{PA}[PA]{Power Amplifier}
\acrodef{RS}[RS]{Resource Sharing}
\acrodef{PC}[PC]{Power Control}
\acrodef{SOTA}[SotA]{State-Of-The-Art}
\acrodef{EE}[EE]{Energy Efficiency}
\acrodef{SINR}[SINR]{Signal-to-Interference-and-Noise-Ratio}
\acrodef{LTE}[LTE]{Long Term Evolution}
\acrodef{EARTH}[EARTH]{Energy Aware Radio and neTwork tecHnologies}
\acrodef{MIMO}[MIMO]{Multiple-Input and Multiple-Output (transmission)}
\acrodef{SISO}[SISO]{Single-Input and Single-Output (transmission)}
\acrodef{RE}[RE]{Resource Element}
\acrodef{SNR}[SNR]{Signal-to-Noise-Ratio}
\acrodef{ACLR}[ACLR]{Adjacent Carrier Leakage Ratio}
\acrodef{PRAIS}[PRAIS]{Power and Resource Allocation Including Sleep}

\pdfoutput=1 

\begin{document}

\title{On Minimizing Base Station Power Consumption}

\author{{Hauke~Holtkamp, Gunther~Auer}\vspace{3mm}\\
DOCOMO Euro-Labs\\
D-80687 Munich, Germany \\Email:
\{holtkamp, auer\}@docomolab-euro.com
\and Harald~Haas\vspace{2mm}\\
Institute for Digital Communications\\
Joint Research Institute for Signal and Image Processing\\
 The University of Edinburgh,
EH9 3JL, Edinburgh, UK\\ E-mail: h.haas@ed.ac.uk}


\maketitle

\begin{abstract}
We consider resource allocation over a wireless downlink where \ac{BS} power consumption is minimized while upholding a set of required link rates. A \ac{PRAIS} method is proposed that combines resource sharing, \ac{PC}, and \ac{DTX}, such that downlink power consumption is minimized. Unlike conventional approaches that aim at minimizing transmit power, in this work the \ac{BS} mains supply power is chosen as the relevant metric. Based on a linear power model, which maps a certain transmit power to the necessary mains supply power, we quantify the fundamental limits of \ac{PRAIS} in terms of achievable \ac{BS} power savings. The fundamental limits are numerically evaluated on link level for four sets of \ac{BS} power model parameters representative of envisaged future hardware developments. For BSs installed around 2014, \ac{PRAIS} provides 63\% to 34\% energy savings over conventional resource allocation schemes, depending on the rate target per link.
\end{abstract}


\acresetall

\section{Introduction}
Recent surveys on the energy consumption of cellular network components throughout the whole life cycle, including \ac{BS}, mobile terminals and core network, reveal that around 80\% of the electricity bill of a mobile network operator are generated at \ac{BS} sites~\cite{fmbf1001, aghimfbhz1001}. 
This highlights the need to improve the energy efficiency at the \ac{BS}.

Moreover, analyzing the traffic data of mobile operators suggests that only during few hours per day \acp{BS} are running under full load --- for which they were designed --- while operating in low load for the rest of the time \cite{aghimfbhz1001,s1001}. This creates potential for \ac{BS} energy savings by tailoring the resource and power allocation for low load situations. How
In the past, energy efficiency research on wireless transmitters has focused on the minimization of transmit power. The EARTH\footnote{EU funded research project EARTH (Energy Aware Radio and neTwork tecHnologies), FP7-ICT-2009-4-247733-EARTH, Jan.~2010 to June~2012. https://www.ict-earth.eu} power model establishes a relation between a given transmit power and the necessary input supply power of a \ac{BS}~\cite{czbfjgav1001}. Two fundamental characteristics of a \ac{BS} are reported: first, the power model is closely approximated by a linear function; and second, the power consumption in idle mode (when no data is transmitted) may be significantly reduced with respect to the active transmission mode, since some hardware components and circuits may be switched-off. This linear power model allows to minimize the \emph{overall} \ac{BS} supply power consumption, rather than transmit power only.


%

\begin{figure}
\centering
\includegraphics[width=0.6\textwidth]{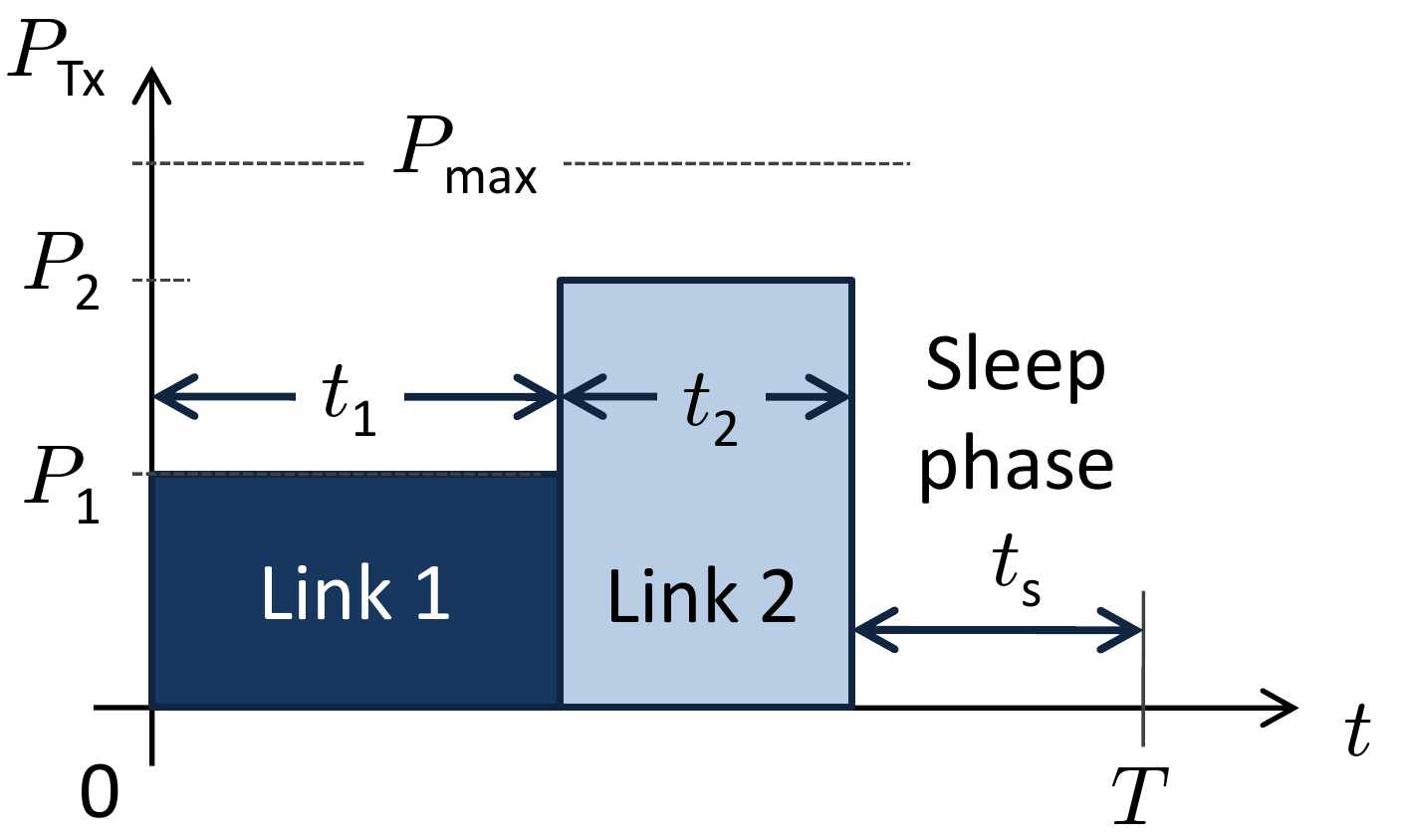}
\caption{Illustration of \ac{PRAIS} for two links: transmit power, resource share and sleep mode are allocated optimally in order to serve the requested rate at minimal overall power consumption.}
\label{fig:PRAIS}
\end{figure}

In this study, the EARTH power model is utilized to extract fundamental limits on efficient energy resource allocation. 
The proposed \ac{PRAIS} scheme combines three techniques: \ac{PC}, appropriate resource sharing between multiple users by means of \ac{TDMA} and \ac{DTX}. See Figure~\ref{fig:PRAIS} for an illustration on two links. \ac{DTX} refers to the ability to put the \ac{BS} into a sleep mode, which has lower consumption than the active or idle state.  To date, \ac{DTX} or \ac{PC} have only been considered individually~\cite{fmmjg1101, ssha0801a}, thus missing the combined gains. The contribution of this work is to optimize \ac{PC}, \ac{TDMA} and \ac{DTX} jointly, such that the \ac{BS} supply power is minimized. For some representative power models that reflect anticipated developments in \ac{BS} hardware, numerical results for the expected power consumption of future \acp{BS} are provided. 

The remainder of this paper is structured as follows. Section~\ref{powermodel} introduces the EARTH power consumption model. Section~\ref{systemmodel} derives the \ac{PRAIS} scheme and the resulting optimization problem and states the scenario parameters. Section~\ref{results} presents the quantitative results. The paper is concluded in Section~\ref{conclusion}.

%

\section{Power Model}
\label{powermodel}

\begin{figure}
\centering
\includegraphics[width=0.6\textwidth]{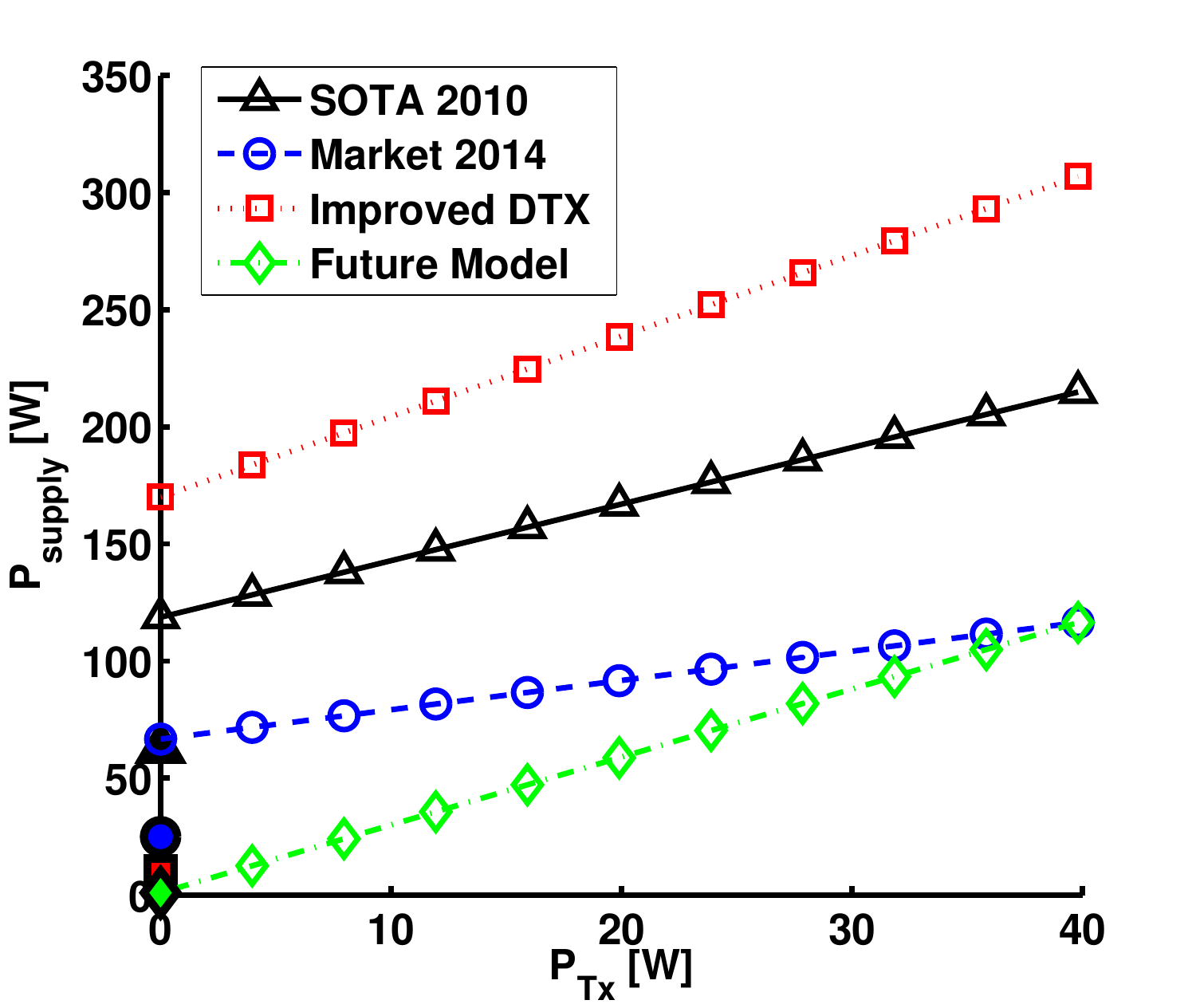}
\caption{Load behavior of the selected power models. Sleep mode consumption $P_s$ is marked on the y-axis by the respective marker.}
\label{fig:powermodels}
\end{figure}

The basis of our analysis is the \ac{BS} power model. A typical \ac{LTE} \ac{BS} consists of such components as radio transceiver, baseband interface, power amplifier, AC-DC-converter, DC-DC-converter and cooling fans, each possibly manifold depending on the number of sectors and antennas employed. Each of these components has an individual power consumption that may either be constant or load-dependent. It was found in~\cite{czbfjgav1001} that the power consumption of such a \ac{SOTA} \ac{BS} can be approximated by a linear function. This reflects the fact that some components have constant consumption independent of the load and the sum of load-dependent components creates an affine function, resulting in a straight line plot with a y-axis offset. 
Analytically, the linear power model for the instantaneous power consumption as a function of the transmit power~$P_{\rm{Tx}}$ can be written as
\begin{equation}
\label{eq:psupply}
 P_{\rm{supply}} = 
  \begin{cases}
   P_0 + m P_{\rm{Tx}} 	& \text{if } 0 < P_{\rm{Tx}} \leq P_{\rm{max}}  \\
   P_{\rm{s}}      	& \text{if } P_{\rm{Tx}} = 0.
  \end{cases}
\end{equation}
where $m$ denotes the slope of the trajectory that quantifies the load dependence, $P_{\rm{max}}$ is the maximum transmit power, $P_0$ and $P_{\rm{s}}$ account for the stand-by and sleep mode consumption of the \ac{BS}, respectively.
The power models used in this study are depicted in Figure~\ref{fig:powermodels} as functions of the transmit power~$P_{\rm{Tx}}$. Also shown in Figure~\ref{fig:powermodels} is the consumption in sleep mode~$P_{\rm{s}}$ as markers on the y-axis for $P_{\rm{Tx}} {=} 0$.

\begin{table}
      \caption{Power Model Parameters}
\centering
      \begin{tabular}{c|ccc}
	Power Model			& $P_0$ /W & $m$ & $P_{\rm{s}}$ /W\\
	\hline
	SOTA 2010			& 119	& 2.4	& 63 \\
	Market 2014			& 67	& 1.25	& 25 \\
	Improved DTX			& 170	& 3.4	& 25 \\
	Future Model			& 1	& 2.9	& 1 \\	
	\end{tabular}
\label{tab:powermodelparameters}
\end{table}

\begin{figure*}[!!t]
\normalsize
\setcounter{MYtempeqncnt}{\value{equation}}
\setcounter{equation}{8}
\begin{equation}
\begin{aligned}
& \underset{\mu, \nu}{\text{minimize}}
& & \overline{P}_{\rm{supply}}(\overline{R}_i) =  \left[ \sum_{i=1}^{N_{\rm{L}}} \mu_i \left( P_0 + m \frac{N_i}{G_i} \left( 2^\frac{\overline{R}_i}{W_i \mu_i} - 1 \right) \right) \right] + \nu P_{\rm{s}} \\
& \text{subject to}
& & \sum_i \mu_i + \nu = 1 \,,\quad
\nu \ge 0 \,,\quad
\mu_i \ge 0\ \forall\,i \\
&&& 0 \leq P_i = \frac{N_i}{G_i} \left( 2^{\frac{\overline{R}_i}{W_i \mu_i}} - 1 \right) \leq P_{\rm{max}}
\end{aligned}
\label{eq:OP}
\end{equation}
\setcounter{equation}{\value{MYtempeqncnt}}
\hrulefill
\end{figure*}

Note that in this study, 'load' refers to the power density in an \ac{OFDMA} system like \ac{LTE}. High load signifies that many data \acp{RE} in a particular time slot are used for transmission, each adding to the overall transmit power. Full load refers to all \acp{RE} in use, leading to highest allowed transmit power.

Representative of current and future developments, we select four different power models; three from literature and a best-case assumption with the parameters listed in Table~\ref{tab:powermodelparameters}. The first set of parameters represents the \ac{SOTA} as of 2010. Power consumption is reported to be $P_{\rm{supply}}=1292$\,W and $712$\,W at full and zero load. Since the \ac{BS} studied in~\cite{aggsoisgd1101} covers three sectors with two transmit antennas each, and the consumption of a \ac{BS} scales approximately linearly with the number of transmit antennas and sectors, we scale down $P_{\rm{supply}}$ by a factor of six to represent a single antenna, one-sector \ac{BS}. 

The second model is a prediction of developments that will available on the market in 2014 \cite{czbfjgav1001}. It is based on anticipated component improvements over four years. We label this the 2014 Market model. 

Third, for the benefit of sleep modes, we follow an assumption where sleep modes will greatly improve, while standby component consumption cannot be reduced substantially in the coming years~\cite{fmmjg1101}. As this model assumes \ac{BS} from some years before 2010, the load-dependent power consumption is higher than in the \ac{SOTA} 2010 model, thus particularly emphasizing \ac{DTX} effects.

As a best-case example, the fourth power model assumes idealized components that scale perfectly with load. Power consumption is set to scale linearly with load with near-zero stand-by consumption. This model is not a prediction of technology advances, but provides theoretical limits. Assuming that operating efficiency at full load will not change after 2014, the future model has consumption equal to the 2014 Market model at full load/maximum transmit power.

Knowledge of the dependence of supply power on transmit power allows the targeted optimization via the \ac{PRAIS} scheme.

\section{PRAIS}
\label{systemmodel}
When employing \ac{DTX} individually to minimize power consumption, all required links are served at the maximum allowed transmission power until target rates are fulfilled, then the \ac{BS} may go to sleep mode. Alternatively, when employing \ac{PC} with \ac{TDMA} individually for maximum efficiency, the transmission duration is stretched over the available time frame with the lowest power necessary to serve the required rates. When \ac{PC} and \ac{DTX} are combined, there is a trade-off. Between the two extremes of maximum transmit power with longest sleep mode and lowest transmit power with no sleep mode, there is a configuration with medium transmit power and medium sleep mode duration that consumes less overall power. This is exploited in the \ac{PRAIS} scheme.

We proceed to derive the power consumption optimization problem for a multi-user single-cell allocation of transmit powers, sleep times and transmit durations contained within the \ac{PRAIS} scheme. We define the normalized duration 
\begin{equation}
 \label{eq:mu}
  \mu_i = \frac{t_i}{T}
\end{equation}
where $t_i$ is the time allocated to link $i$ and $T$ is the duration of the considered time frame in seconds.

The normalized duration spent in sleep mode is
\begin{equation}
 \label{eq:nu}
 \nu = \frac{t_{\rm{s}}}{T}
\end{equation}
where $t_{\rm{s}}$ is the time spent in sleep mode and 
\begin{equation}
 \displaystyle\sum_{i=1}^{N_{\rm{L}}} \mu_i + \nu = 1 
\end{equation}
with $N_{\rm{L}}$ the number of links served.

Since we are interested in the fundamental limits we employ the Shannon bound to associate transmit powers with link rates. The rate on each link is  
\begin{equation}
 R_i(P_i) = W_i \rm{log}_2 ( 1 + \gamma_i \it{(P_i)}),
\end{equation}
where $\gamma_i(P_i)=\frac{G_i P_i}{N_i}$ denotes the \ac{SNR} and $W_i$ is the transmission bandwidth in Hz, with channel gain $G_i$, transmit power $P_i{\le}P_{\rm{max}}$ in W. The noise power is defined by $N_i = W_i k \vartheta$ with Boltzmann constant~$k$ and operating temperature~$\vartheta$ in Kelvin. 

The average rate on link~$i$ over the time slot~$T$ is 
\begin{equation}
 \overline{R}_i = \mu_i R_i.
\end{equation}

Thus, the average transmit power required to fulfill a target average rate $\overline{R}_i$ is
\begin{equation}
\label{eq:ptxavg}
 \overline{P}_{\rm{Tx},\it{i}}(\overline{R}_i) = \frac{N_i}{G_i} \left( 2^\frac{\overline{R}_i}{W_i \mu_i} - 1 \right).
\end{equation}

The overall power consumption caused when transmitting is the weighted average of the link supply power consumptions found by summation of \eqref{eq:psupply} over all links: 
\begin{equation}
 \overline{P}_{\rm{active}}(\overline{R}_i)  = \sum_{i=1}^{N_{\rm{L}}} \mu_i \left( P_0 + m P_{\rm{Tx},i}(\overline{R}_i) \right).
\end{equation}

Combining with the consumption during sleep mode from \eqref{eq:psupply} and \eqref{eq:ptxavg} we generate the cost function for $P_{\rm{supply}}$ yielding the optimization problem~\eqref{eq:OP}, displayed at the top of this page. The constraints reflect the facts that normalized time has to be positive and their sum unity. Transmission powers are positive and bounded by a maximum transmit power. The cost function is a non-negative sum of functions that is convex within the constraint domain and is therefore still convex. See the appendix for a proof. It can be solved with appropriate software like the MATLAB optimization toolbox. The solution of \eqref{eq:OP} determines the vectors $\mu$ and $\nu$ which minimize the overall power consumption under target rates. Note that fixating $P_0 = P_{\rm{s}}$ is equivalent to disabling \ac{DTX}, \ie\ employing \ac{PC} individually, whereas fixating $P_{\rm{Tx}} = P_{\rm{max}}$ is equivalent to disabling \ac{PC}. 

The resource allocation problem~\eqref{eq:OP} can be solved for any number of users. Without loss of generality, we select a ten user scenario for numerical evaluation.


\section{Results}
\label{results}

\begin{table}
      \caption{Simulation Parameters}
\centering
      \begin{tabular}{c|c}
	Parameter          			& Value\\
	\hline
	Carrier frequency  			& 2\,GHz\\
	Cell radius  				& 250\,m\\
	Pathloss model  			& 3GPP UMa, NLOS, shadowing~\cite{std:3gpp-faeutrapla}\\
	Shadowing standard deviation		& 8\,dB\\
	Iterations  				& 10,000 \\
	Bandwidth $W$ 				& 10\,MHz \\
	Maximum transmission power $P_{\rm{max}}$	& 46\,dBm \\
	Operating temperature $T$ 		& 290\,K \\
	\end{tabular}
\label{tab:uniformdistrscenarioparameters}
\end{table}

For the numerical analysis, we evaluate the \ac{PRAIS} scheme in a Monte Carlo simulation under the parameters shown in Table~\ref{tab:uniformdistrscenarioparameters}. Users are dropped uniformly onto a disk and the associated channel gains~$G_i$ are generated applying the 3GPP urban macro path-loss model including shadowing with a standard deviation of~$8$\,dB. In addition to the individual \ac{DTX} and \ac{PC} allocation schemes and the \ac{PRAIS} scheme, we present two references that serve as upper limits. First, the maximum transmission power as defined by the \ac{LTE} standard provides the theoretical reference. Second, we set the reference against which gains are measured to be the power behavior of a \ac{BS} as defined in the power models. Practically, this is a \ac{FDMA} scheme where each user receives a share of the frequency band as well as the entire considered time slot. This can be interpreted as '\ac{DTX} in the frequency domain' where $P_{\rm{s}} = P_0$. In essence, this is bandwidth adaptation and is our reference against which we assess the achieved savings. 

\begin{figure*}[!t]
\centering
\subfigure[Supply power consumption under the SOTA 2010 model.]{
\includegraphics[width=0.5\textwidth]{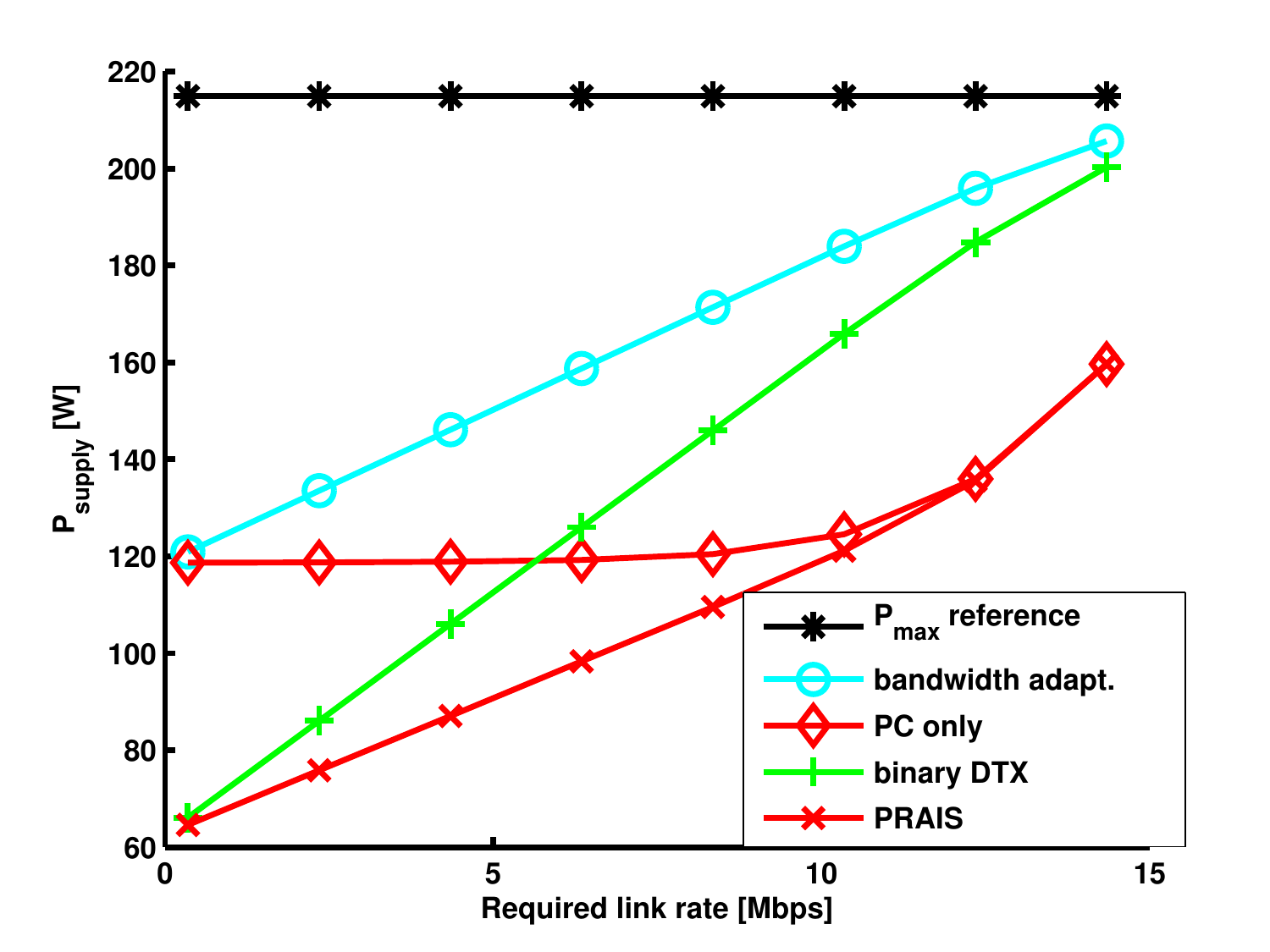}
\label{fig:resultsPM1}
}
\subfigure[Supply power consumption under the Market 2014 model.]{
\includegraphics[width=0.5\textwidth]{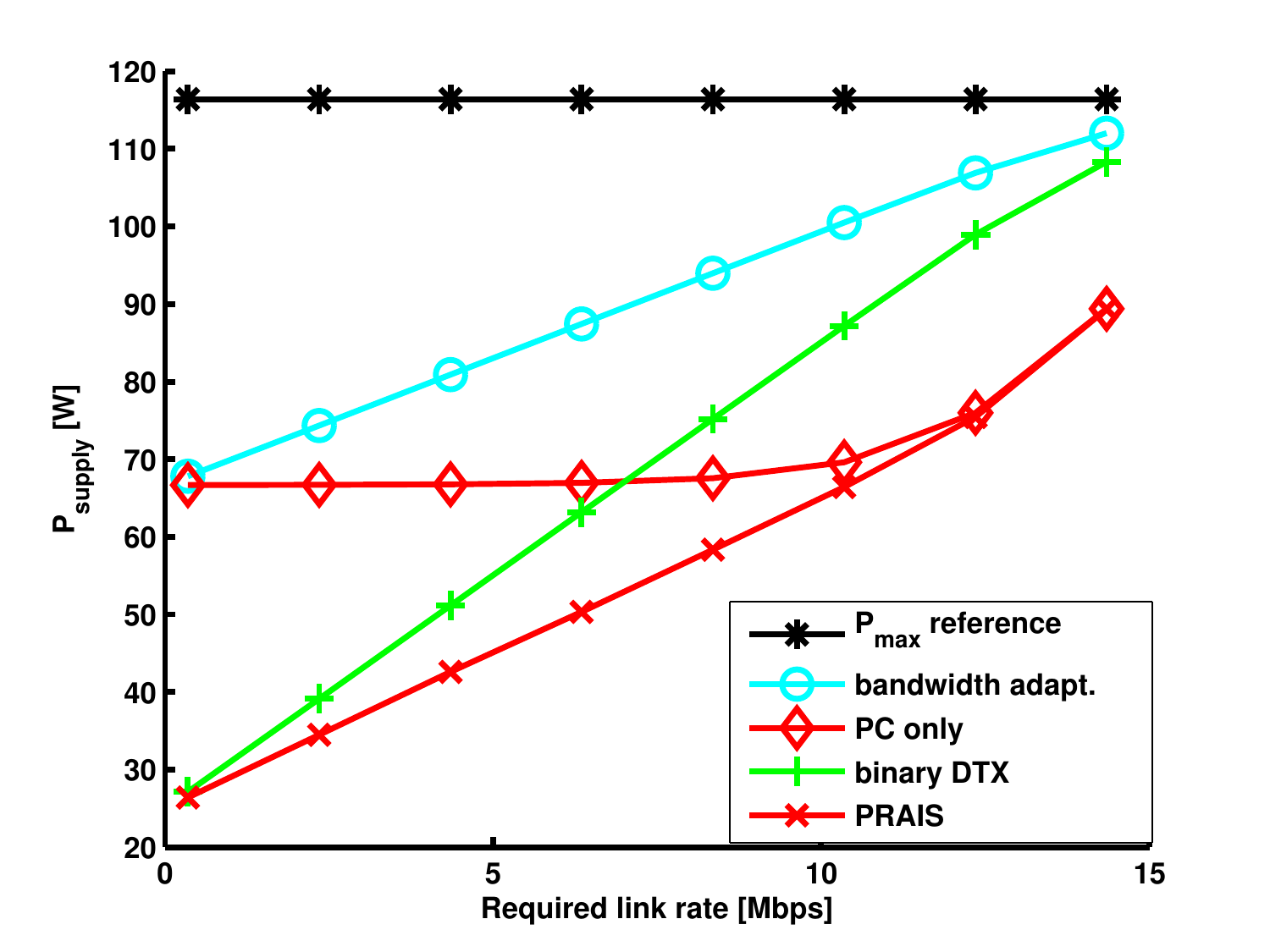}
\label{fig:resultsPM2}
}
\subfigure[Supply power consumption under the Improved DTX model.]{
\includegraphics[width=0.5\textwidth]{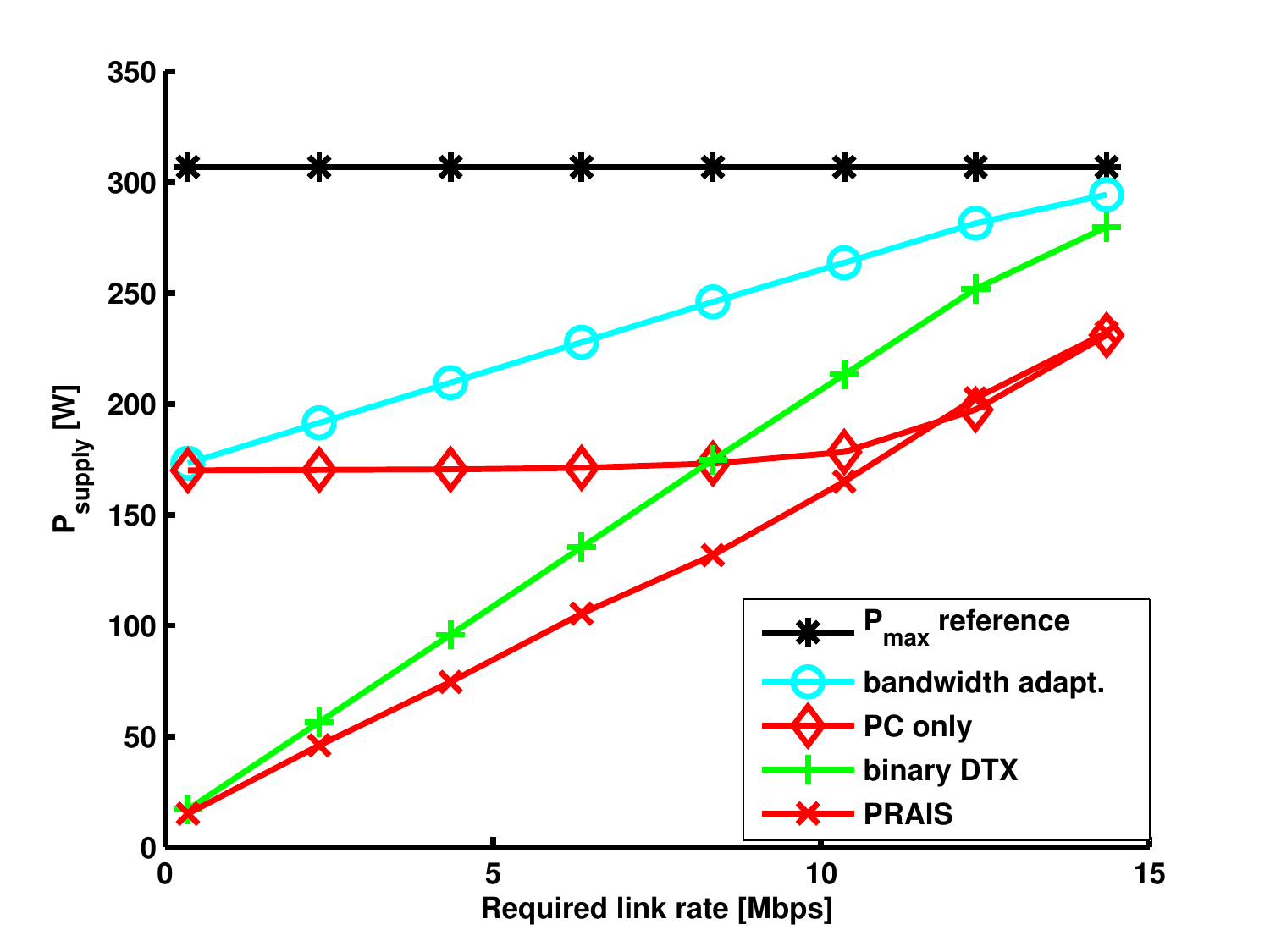}
\label{fig:resultsPM3}
}
\subfigure[Supply power consumption under the Future Model.]{
\includegraphics[width=0.5\textwidth]{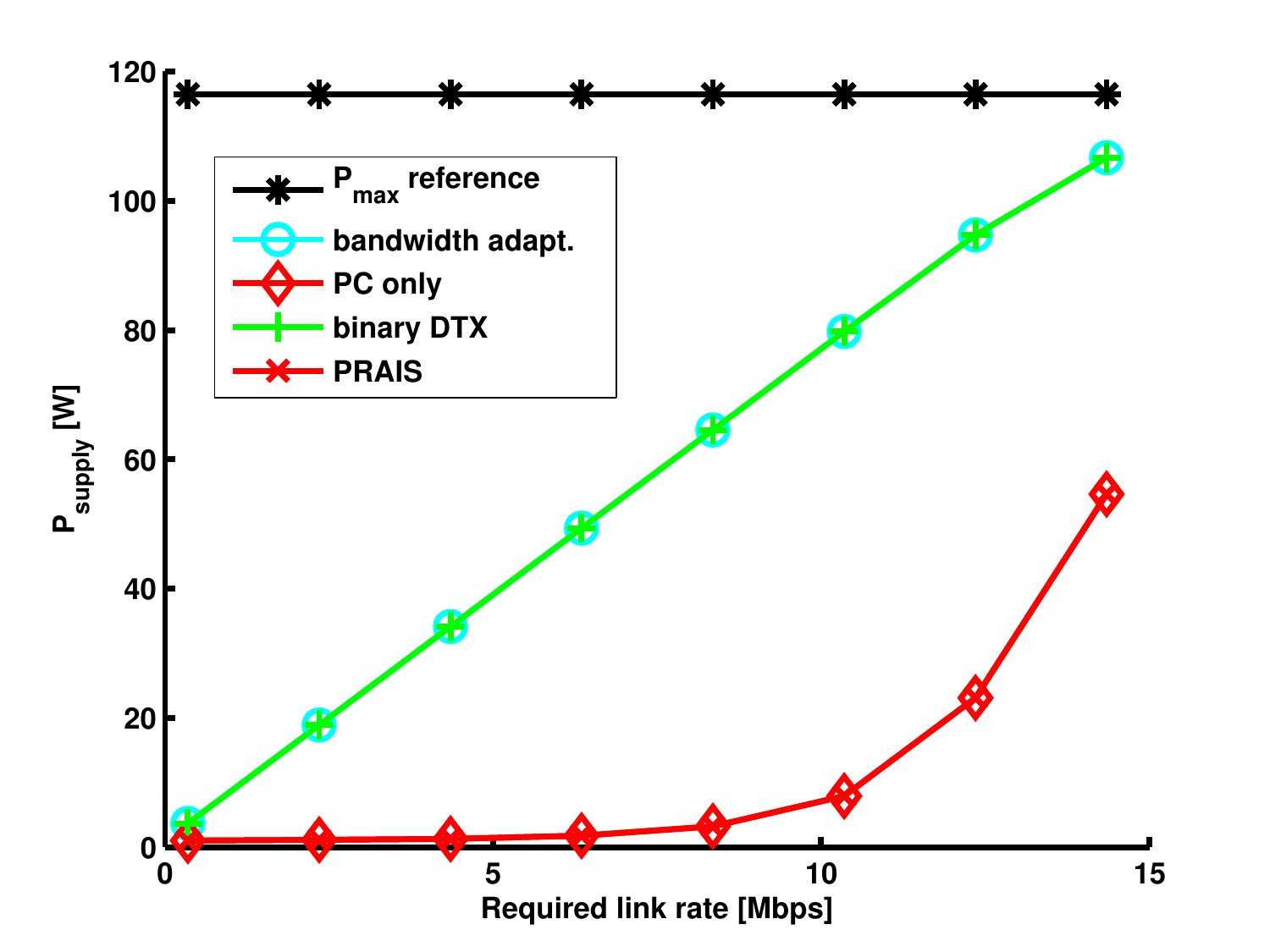}
\label{fig:resultsPM4}
}
\caption{Fundamental limits for power consumption in \acp{BS}.\label{fig:results}}
\end{figure*}

The simulation results are shown in Figure~\ref{fig:results} where the per-user link rate is plotted against the average supply power consumption under the different schemes. Figures ~\ref{fig:resultsPM1}, \ref{fig:resultsPM2}, \ref{fig:resultsPM3}, \ref{fig:resultsPM4} reflect the four chosen power models. 

In Figure~\ref{fig:resultsPM1} we see that a \ac{SISO} link in an \ac{LTE} \ac{BS} of 2010 consumes 120\,W up to 220\,W in bandwidth adaptation, which we consider the operation of the \ac{SOTA}. The first result is that the consumption curves of bandwidth adaptation and \ac{PC} as well as \ac{DTX} and \ac{PRAIS} originate in the same values of $P_{\rm{supply}}$. The higher one is $P_0$ at 119~W and the lower one $P_{\rm{s}}$ at 67~W. 

The use of \ac{PC} only allows to keep the overall consumption constant for a large set of low target rates. In this rate region, transmit powers are very low compared to standby consumption. Only when rates above 10~Mbps are requested is the transmission power high enough to make a noticeable difference compared to the standby consumption. This is reflected in the rising \ac{PC} only curve above 10~Mbps. In contrast, the binary \ac{DTX} scheme has much lower power consumption (up to $-45$\%) at low rates. However, it rises much quicker than the \ac{PC} only curve. There is a crossover point at 5.6~Mbps between \ac{DTX} and \ac{PC}. The combined \ac{PRAIS} scheme has minimal consumption over all rates. At higher target rates, the \ac{PRAIS} consumption curve joins the \ac{PC} only curve. This reflects the fact, that it is not feasible to put the \ac{BS} to sleep at high rates. In the 2010 model, the \ac{PRAIS} scheme achieves savings between 45\% and 31\% over bandwidth adaptation.

Similar behavior can be found in the 2014 Market model in Figure~\ref{fig:resultsPM2}. The model predicts that sleep mode consumption will be less than half of the standby consumption at 25\,W. This favors the application of \ac{DTX} which is lower bounded by this value. The cross-over point between the individual \ac{DTX} and {PC} schemes shifts to the right to around 7~Mbps. This means that \ac{DTX} is more efficient than \ac{PC} for a larger set of rates. We see this as a trend, especially since sleep modes in \acp{BS} are a new trend which will receive strong research efforts on the hardware side. In the 2014 model, \ac{PRAIS} delivers savings between 61\% and 34\% over bandwidth adaptation.

A surprising result is found in Figure~\ref{fig:resultsPM3}. Although this model has a strong bias towards \ac{DTX} effects, there remains a cross-over point between \ac{DTX} and \ac{PC} after which the use of \ac{PC} can still add an extra 15\% savings on top of \ac{DTX}. In the Improved DTX model, the \ac{PRAIS} scheme offers between 90\% savings at near-zero rates and 38\% savings at 10~Mbps.

In the future linear model in Figure~\ref{fig:resultsPM4} the behavior of \ac{DTX} and bandwidth adaptation, as well as \ac{PC} and \ac{PRAIS}, are identical, due to the fact that there is no gain of sleep modes over standby consumption. Bandwidth adaptation consumption is significantly lower than for all other power models. All gains are realized by \ac{PC} which is strongly amplified by the linear behavior, delivering gains of up to 75\% over the bandwidth allocation scheme.  

With regard to fundamental limits we find that highest consumption and thus the upper limit of the downlink power consumption is found to be a 2010 \ac{BS} without energy efficient allocation at 119--220\,W. When upscaling by three sectors and two transmit radio chains, we arrive at a consumption of 712\,W to 1121\,W. For the year 2014, we assume the default configuration of \acp{BS} to be equipped with four transmit antennas and full application of \ac{PRAIS}. Thus, the consumption of 25--71\,W must be upscaled by a factor of 12, resulting in an expected consumption of 300\,W to 852\,W per \ac{BS} in 2014. This is an important finding, because it states that although \acp{BS} will contain more radio chains in the future, power consumption can still be expected to decrease (if energy efficiency measures are applied). 

The attainable power savings of PC, DTX and PRAIS hardly depend on the number of users; gains decrease slightly as the number of users increases (not shown in Fig.~\ref{fig:results}).


\section{Conclusion}
\label{conclusion}
In this paper we have presented a comparison of \ac{TDMA}, \ac{PC}, \ac{DTX} and bandwidth adaptation individually in the cellular downlink on the power supply metric. This leads to the optimal \ac{PRAIS} scheme which exploits \ac{TDMA}, \ac{PC} and \ac{DTX} in combination. It was found that the \ac{PRAIS} scheme is convex which allows to find a global minimum of the supply power cost function. It can be seen in simulation that \ac{DTX} contributes the highest savings compared to the other schemes. The \ac{PRAIS} scheme can lower the power consumption of the 2014 Market \ac{BS} by between 61\% and 34\%.

We find that from the two investigated individual schemes, \ac{DTX} provides higher individual gains under the assumptions made in this study. But even under extreme assumptions that strongly benefit the use of \ac{DTX}, the combined use of \ac{DTX} and \ac{PC} within the \ac{PRAIS} scheme still delivers significant gains over the individual scheme. The two power saving techniques of \ac{PC} and \ac{DTX} are shown to have different applicability with \ac{DTX} in the low rates and \ac{PC} in high rates. 

As a result of the overall predicted consumption development, we consider the adaptation of the number of active radio chains to the load an important future research topic.

Since practical limitations may reduce the applicability of \ac{PRAIS} our future work will take more limitations into account, \eg~like \ac{LTE} control signalling. 

Ultimately, we establish 27\,W at near-zero load and 68\,W at 10\,Mbps per 10\,MHz bandwidth and 10 users as the lower fundamental limit of a \ac{BS} in 2014 when employing resource and power allocation via \ac{PRAIS}. This represents a saving over bandwidth allocation of 45\% to 31\% in 2010 and 61\% to 34\% in 2014, respectively.

\setcounter{equation}{9}

\appendix
We show that \eqref{eq:OP} is monotonically decreasing and convex without sleep modes and that it is still convex (but no longer monotonically decreasing) when sleep modes are considered. The first derivative of the cost function for link~$i$ is
\begin{equation}
\begin{aligned}
 f_{0,i}^\prime & = \frac{N_i}{G_i} \left[ - 1 + 2^\frac{\overline{R_i}}{W_i \mu_i} \left( 1 - \frac{1}{\mu_i} \frac{\overline{R_i}}{W_i} \ln(2) \right) \right].
\end{aligned}
\label{eq:f0iprimePC}
\end{equation}
It is seen that for $\mu_i \rightarrow 0$ the negative term goes towards infinity, thus $f_{0,i}^\prime \rightarrow -\infty$. For $\mu_i \rightarrow 1$, the outcome depends on $\frac{\overline{R_i}}{W_i}$. 
For $\frac{\overline{R_i}}{W_i} \rightarrow +0$, $f_{0,i}^\prime \rightarrow -0$. 
For $\frac{\overline{R_i}}{W_i} \rightarrow +\infty$, $f_{0,i}^\prime \rightarrow -\infty$. 
In all cases, the first derivative is negative, thus the cost function is monotonically decreasing in $\mu_i$. 
The second derivative of~$f_{0,i}$ is given by
\begin{equation}
 f_{0,i}^{\prime\prime} = \frac{N_i}{G_i} \left(\frac{\overline{R_i}}{W_i} \right)^2 \ln^2(2) \frac{ 2^\frac{\overline{R_i}}{W_i \mu_i}}{\mu_i^3}
\label{eq:f0iprimeprimePC}
\end{equation}
All variables in the second derivative are positive, thus $f_{0,i}^{\prime\prime} \geq 0$ within the parameter bounds. Therefore, each $f_{0,i}$ is convex within the bounds. 

The non-negative sum preserves convexity. Thus, $f_{0}$ is convex. 

When $P_{\rm{s}} > 0$, the first derivative is no longer negative for all $\mu$, thus the cost function is no longer monotonically decreasing in $\mu_i$. However, the linear terms drops out in the second derivative, such that only the second derivative \eqref{eq:f0iprimeprimePC} remains, which has been shown to be convex.

\section*{Acknowledgements}

This work has received funding from the European Community's $7^{\textrm{th}}$ Framework Programme [FP7/2007-2013. EARTH, Energy Aware Radio and neTwork tecHnologies] under grant agreement n${}^\circ$ 247733.

The authors gratefully acknowledge the invaluable insights and visions received from partners of the EARTH consortium.


\bibliography{../../../../../../DOCOMO/reference/DOCOMO,../../../../../../DOCOMO/reference/general}

\bibliographystyle{ieeetr}

\end{document}